\documentclass{article}
\usepackage{dsfont,amscd,amsmath,amssymb}
\def\DH{\rm I\kern-1.5pt\rm H\kern-1.5pt\rm I}

\newcommand{\mq}{\mathbf{q}}
\newcommand{\mpsi}{\boldsymbol{\psi}}
\newcommand{\mbpsi}{\bar{\boldsymbol{\psi}}}
\newcommand{\mlambda}{\boldsymbol{\lambda}}
\newcommand{\mblambda}{\bar{\boldsymbol{\lambda}}}
\newcommand{\blambda}{\bar{\lambda}}
\newcommand{\mLambda}{\boldsymbol{\Lambda}}
\newcommand{\mbLambda}{\overline{\boldsymbol{\Lambda}}}

\newcommand{\p}[1]{(\ref{#1})}

\newcommand{\cD}{{\cal D}}

\newcommand{\cH}{{\cal H}}

\newcommand{\bT}{{\overline T}{}}
\newcommand{\bZ}{{\overline Z}{}}

\newcommand{\bD}{{\overline D}{}}

\newcommand{\bQ}{{\overline Q}{}}
\newcommand{\bS}{{\overline S}{}}
\newcommand{\bLambda}{{\overline \Lambda}{}}

\newcommand{\bpsi}{{\bar\psi}{}}

\newcommand{\bq}{{\bar q}}
\newcommand{\bnabla}{{\bar \nabla}}
\newcommand{\nablath}{{\nabla_\theta}}
\newcommand{\bnablath}{{\bar \nabla_\theta}}

\newcommand{\cE}{{ {\cal E}   }}
\newcommand{\be}{\begin{equation}}
\newcommand{\ee}{\end{equation}}
\newcommand{\bea}{\begin{eqnarray}}
\newcommand{\eea}{\end{eqnarray}}
\newcommand{\und}{\qquad\textrm{and}\qquad}

\newcommand{\ba}{\begin{array}} \newcommand{\ea}{\end{array}}

\def\im{{\rm i}}

\newcommand{\nn}{\nonumber}
\usepackage{amscd,amsmath,amssymb}

\begin{document}
\thispagestyle{empty}
\begin{flushright}
ITP-UH-19/13 \\
\end{flushright}\vspace{1.3cm}
\begin{center}
{\LARGE\bf Higher-derivative ${\cal N}{=}\,4$ superparticle \\[3mm]
in three-dimensional spacetime}
\end{center}
\vspace{2cm}
\begin{center}
{\Large Nikolay~Kozyrev$\;^a$, Sergey~Krivonos$\;^{a}$,
Olaf~Lechtenfeld$\;^{b}$, Armen~Nersessian$\;^{c}$ }
\end{center}
\vspace{1cm}
\begin{center}
$\;^a$ {\sl Bogoliubov Laboratory of Theoretical Physics,
Joint Institute for Nuclear Research, Dubna, Russia }\\[8pt]
${}^b$ {\sl Leibniz Universit\"at Hannover,
Appelstrasse 2, 30167 Hannover, Germany}\\[8pt]
$\;^c${\sl Yerevan State University, 1 Alex
Manoogian St., Yerevan, 0025, Armenia}
\end{center}
\vspace{5cm}
\begin{abstract}\noindent
Using the coset approach (nonlinear realization) we construct component actions for a superparticle
in three-dimensional spacetime with ${\cal N}{=}\,4$ supersymmetry partially broken to ${\cal N}{=}\,2$.
These actions may contain an anyonic term and the square of the first extrinsic worldline curvature.
We present the supercharges for the unbroken and broken supersymmetries as well as the Hamiltonian
for the supersymmetric anyon. In  terms of the nonlinear realization superfields, the superspace actions
take a simple form in all cases.
\end{abstract}

\newpage

\setcounter{page}{1}
\setcounter{equation}{0}
\section{Introduction}
In a relativistic theory, any particle-like configuration spontaneously breaks the target-space
Poincar\'{e} invariance to the stability group of the worldline.
This breakdown is accompanied by the appearing of Goldstone bosons associated with
the spontaneously broken translations and Lorentz boosts. The most appropriate method to construct
low-energy effective actions for these Goldstone modes is the nonlinear-realization (or coset) approach
\cite{coset1}, suitably modified for the case of supersymmetric spacetime symmetries~\cite{coset2}.

Towards the construction of particle actions in $D$-dimensional spacetime, the coset approach works as follows.
Let $\{P, Z_i\}$ with $i=1,\ldots,D{-}1$ be the generators of the target spacetime  translations,
$\{M_{ij}\}$ be the generators of the $SO(D{-}1)$ subgroup of the Lorentz group $SO(1,D{-}1)$
rotating the spatial coordinates $Z_i$ among each other, and $\{K_i\}$ generate the coset $SO(1,D{-}1)/SO(D{-}1)$.
All transformations of the full Poincar\'{e} group may be realized by a left action on the coset element
\be\label{I1}
g= e^{t P} e^{q_i (t) Z_i} e^{\Lambda_i(t)K_i}.
\ee
The dependence of the coset coordinates $q_i(t)$ and $\Lambda_i(t)$ on the time $t$ signals
that the $Z$ and $K$ symmetries are spontaneously broken.

According to the general theorem \cite{ih}, not all of the above Goldstone fields have to be treated as independent.
In the present case, the fields $\Lambda_i(t)$ can be covariantly expressed through time derivatives of $q_i(t)$
by imposing the constraint
\be\label{I3}
\Omega^i_Z=0,
\ee
where the Cartan forms $\Omega$ are defined in a standard way,
\be\label{I2}
g^{-1}d g = \Omega_P P + \Omega^{ij}_M M_{ij}+ \Omega^i_Z Z_i+\Omega^i_K K_i.
\ee
Thus, we are dealing with the fields $q_i(t)$ only. The form $\Omega_P$ defines the einbein $E$,
which connects the covariant world-volume form $\Omega_P$ and the differential $dt$ via
\be\label{I4e}
\Omega_P = E \; dt .
\ee
Observing that the form $\Omega_P$ is invariant under all symmetries, one
may immediately write an invariant action~\cite{town1,town2, GKP, GKW},
\be\label{I4}
S_0 =  \int dt\; E.
\ee
This action describes a free particle moving in $D$-dimensional spacetime in the static gauge.

The Cartan forms $\Omega^i_K$ pertaining to the coset may be used for constructing actions
with higher time derivatives~\cite{asluk1,ksquare,rigid,town2,GKP}. Moreover, in three spacetime dimensions,
$D{=}3$, there exists an additional possibility: the form $\Omega_M$ allows for a Wess-Zumino-like term
in the action,
\be\label{I5}
S_{\rm anyon} =  \frac{\alpha}{2} \int \Omega_M,
\ee
which provides the system with a non-zero (anyonic) spin~\cite{anyon}. The above integrand $\Omega_M$
is only quasi-invariant under the three-dimensional Poincar\'{e} transformations~\cite{asluk2},
i.e.~it shifts by a full time derivative under $K_i$ transformations.

The supersymmetric generalization of particle actions within the coset approach requires spinor generators
$Q$ and $S$ which extend the Poincar\'{e} group to the super-Poincar\'{e} one:
\be
\left\{ Q,Q\right\} \sim P, \quad \left\{ S,S\right\} \sim P, \quad \left\{ Q,S\right\} \sim Z.
\ee
All symmetries can then be realized by group elements acting on the coset element
\be\label{I6}
g= e^{t P} e^{\theta^a Q_a} e^{q^i(t,\theta) Z_i} e^{\psi^a(t,\theta)S_a} e^{\Lambda^{i}(t,\theta)K_i}.
\ee
One obtains a collection $\{q^i(t,\theta), \psi^a(t,\theta), \Lambda^i(t,\theta)\}$ of Goldstone superfields
which depend on the worldline superspace coordinates~$\{t, \theta\}$.
The rest of the coset approach machinery works as before:
one may construct the Cartan forms $g^{-1}dg$ for the coset element~\p{I6}
(and obtain new forms $\Omega_Q$ and $\Omega_S$),
one may find the supersymmetric einbein and the corresponding bosonic and spinor covariant derivatives
$\nabla_P$ and $\nabla_Q$, respectively.
One may even invent proper generalizations of the covariant constraints \p{I3} as
\be\label{I7}
\Omega_Z=0, \quad \Omega_S|=0,
\ee
where $|$ denotes the $d\theta$-projection of a form (see e.g.~\cite{Iv} and references therein).
The structure of the coset element~\p{I6} implicates that $Q$ supersymmetry is kept unbroken
while $S$ supersymmetry is spontaneously broken.\footnote{
In this paper we shall only consider the case where $\#Q=\#S$, i.e.~a half-breaking of global supersymmetry.}

The constraints \p{I7} leave the lowest components of the superfields $q^i(t,\theta)$ and $\psi^a(t,\theta)$
as the only independent component fields of the theory.
Unfortunately, like it happened in~\p{I5}, any superparticle Lagrangian is only {\it quasi-invariant\/}
with respect to the super-Poincar\'{e} group.
For this reason, the corresponding action cannot be built from the Cartan forms.
Commonly adopted alternatives for constructing supersymmetric particle (or brane) actions are
\begin{itemize}
\item to construct a linear realization of target-space Poincar\'{e} supersymmetry,
in which the superfield Lagrangian appears as a supermultiplet component~\cite{BG2,RT,IK1},
\item to perform a reduction from higher-dimensional component actions,
\item to make a superfield ansatz for the action (manifestly invariant under $Q$ supersymmetry)
and then impose the spontaneously broken $S$ supersymmetry invariance.
\end{itemize}
Clearly, in all these approaches the coset method is not too helpful.
The method working perfectly in bosonic models seems to be almost useless in the supersymmetric case!
This shortcoming is caused by our concentrating on unbroken $Q$ supersymmetry and on the superspace action.
If instead we focus on the component action with broken $S$ supersymmetry being manifest,
the coset approach will again be quite useful. It has indeed been demonstrated in \cite{BKS1,BKKS1} that,
with the coset parametrization~\p{I6}, it is easy to produce an ansatz for the component action
manifestly invariant with respect to the broken $S$ supersymmetry.
To this end, the following properties are important:
\begin{itemize}
\item with the chosen parametrization \p{I6} of the coset element, the superspace coordinates $\theta$
are inert under $S$ supersymmetry. Therefore, all superfield components transform {\it independently\/}
with respect to $S$ supersymmetry,
\item the $\theta{=}0$ projection of the covariant derivative $\nabla_P$ is invariant under the broken
$S$ supersymmetry,
\item all physical fermionic components are just $\theta{=}0$ projections of the superfields $\psi^a(t, \theta)$,
and these components transform as the fermions of the Volkov-Akulov model \cite{VA} with respect to
the broken $S$ supersymmetry.
\end{itemize}
Thus, an ansatz for the component action with the smallest number of time derivatives can be written down
immediately, because  the physical fermionic components can enter the action only through the $\theta{=}0$
projection of the einbein~$E$ or through the spacetime derivatives~$\nabla_P$ of the ``matter fields'' $q^i(t)$.
This ansatz will contain some arbitrary functions which can be determined by two additional requirements:
\begin{itemize}
\item the supersymmetric action should have a proper bosonic limit,
\item the supersymmetric action has to be invariant under unbroken supersymmetry.
\end{itemize}
These conditions completely fix the component action. Actions for $D{=}2{+}1$ superparticles
realizing an ${\cal N}=2^{k+1}\rightarrow {\cal N}=2^{k}$ pattern of supersymmetry breaking
have been constructed in such a way~\cite{BKKS1}.

The situation becomes more interesting if we admit terms with a non-minimal number of time derivatives
in the action. The main goal of the present paper is to demonstrate how the corresponding component actions
can be constructed for a three-dimensional superparticle with ${\cal N}{=}4$ supersymmetry partially broken
to ${\cal N}{=}2$ and how an anyonic term~\p{I5} and the first extrinsic curvature (``rigidity'')
come to appear in the action.
It should be clear from our exposition that the choice of the physical fermionic components is very important:
it is the choice of the coset element as in~\p{I6} which forces the $\psi|_{\theta=0}$ components to be
Volkov-Akulov goldstini.
In terms of these fermions all the actions we will construct have a clear geometric interpretation.
For the super anyonic case we will provide the Hamiltonian description as well.
For completeness, for all cases considered we will also present the superspace actions which,
in terms of the superfields $\{q^i(t,\theta),\psi^a(t,\theta)\}$, take a simple form.
We shall conclude with a few comments and remarks.

\newpage

\setcounter{equation}{0}

\section{Spontaneous breakdown of $D=2{+}1$ Poincar\'{e} symmetry}
\subsection{Coset approach: kinematics}
The commutation relations of the $D=2{+}1$ Poincar\'{e} algebra read
\bea\label{3dalg}
&& \left[ M_{ab}, P_{cd}\right] = \epsilon_{ac} P_{bd}+\epsilon_{bd} P_{ac}+\epsilon_{ad}P_{bc}+\epsilon_{bc}P_{ad}
, \nn \\
&&\left[ M_{ab}, M_{cd}\right] = \epsilon_{ac} M_{bd}+\epsilon_{bd} M_{ac}+\epsilon_{ad}M_{bc}+\epsilon_{bc}M_{ad} .
\eea
To get a convenient $d=1$ form let us define the following generators,
\bea\label{1dN4not}
&& P= P_{11}+P_{22}, \quad Z=P_{11}-P_{22}-2 \im P_{12},\; \bZ=P_{11}-P_{22}+2 \im P_{12}, \nn \\
&& J=\frac{\im}{4}\left( M_{11}+M_{22}\right), \quad T= \frac{\im}{4}\left( M_{11}-M_{22}-2 \im M_{12}\right),\;
\bT= \frac{\im}{4}\left( M_{11}-M_{22}+2 i M_{12}\right) .
\eea
Being rewritten in terms of these generators \p{1dN4not} the algebra \p{3dalg} acquires the familiar $d=1$ form,
\bea\label{N4d1}
&& \left[J, T\right]= T, \; \left[J, \bT\right]= -\bT, \; \left[T, \bT\right]= -2 J , \nn \\
&& \begin{array}{l}
\left[J,Z\right] =Z, \\
\left[ J, \bZ\right]= - \bZ ,
\end{array}
\qquad
\begin{array}{l}
\left[T,P\right] =-Z, \\
\left[ T, \bZ\right]= - 2 P ,
\end{array}
\qquad
\begin{array}{l}
\left[\bT,P\right] =\bZ, \\
\left[ \bT, Z\right]=  2 P ,
\end{array}
\eea
{}From the $d=1$ point of view the generators $\{ Z, \bZ\}$ are the central charges generators.

We are going to consider the spontaneous breakdown of $D=2{+}1$ Poincar\'{e} symmetry down to $d=1$ Poincar\'{e}, generated
by $P$ and $U(1)$ rotations, generating by $J$. Therefore,  we will put the  generator $J$ in the stability subgroup and choose the parametrization of our coset as
\be\label{N4coset}
g= e^{\im t P}\;  e^{\im\left( q Z + {\bar q} \bZ\right)} \; e^{\im\left( \Lambda T + \bLambda \bT\right)}.
\ee
Here, $q(t), {\bar q}(t), \Lambda(t), \bLambda(t)$ are Goldstone fields depending on the time $t$.

The local geometric properties of the system are specified by the left-invariant Cartan forms
\be\label{coset}
g^{-1}dg=\im \omega_P P+ \im \omega_Z Z+\im\bar\omega_Z {\bar Z} +\im \omega_T T+\im \bar\omega_T {\bar T}+\im\omega_J J
\ee
which look extremely simple,
\bea\label{N4CF}
& \omega_P = \frac{1}{1-\lambda\blambda}\left[ \left(1+\lambda\blambda\right) dt + 2 \im \left( \lambda d{\bar q}-\blambda dq\right)\right], &\nn \\[2mm]
& \omega_Z = \frac{1}{1-\lambda\blambda}\left[dq -\lambda^2 d{\bar q} +\im \lambda dt\right],\qquad
{\bar\omega}_Z = \frac{1}{1-\lambda\blambda}\left[d {\bar q} -\blambda^2 dq -\im \blambda dt\right],&  \\[2mm]
& \omega_T = \frac{d\lambda}{1-\lambda\blambda}, \qquad
{\bar\omega}_T = \frac{d\blambda}{1-\lambda\blambda}, \qquad
\omega_J = \im\frac{\lambda d\blambda-d\lambda \blambda}{1-\lambda\blambda},&\nn
\eea
where
\be\label{N4def2}
\lambda = \frac{ \tanh (\sqrt{\Lambda \bLambda})}{\sqrt{\Lambda \bLambda}}\Lambda \und
\blambda = \frac{ \tanh (\sqrt{\Lambda \bLambda})}{\sqrt{\Lambda \bLambda}}\bLambda .
\ee

The transformations properties of the coordinates and fields are induced by the left multiplications of the coset element \p{coset},
\be
g_0 g= g' h,
\ee
where $h \in U(1)$ belong to the stability subgroup. Thus, for the mostly interesting transformations with
$g_0= e^{i\left( \alpha T +\bar\alpha \bT\right)}$ one gets
\be
  \delta t = - 2 \im \left( \alpha {\bar q}-{\bar\alpha}q\right) , \qquad
 \delta q = -\im \alpha\; t,\qquad \delta{\bar  q} = \im \bar\alpha\; t ,\qquad
 \delta\lambda =\alpha-\bar\alpha\lambda{}^2,\; \delta\blambda= \bar\alpha - \alpha\blambda{}^2.
\ee

Finally, one may reduce the number of independent Goldstone fields by imposing the following conditions on the Cartan forms $\omega_Z$ and ${\bar\omega}_Z$ (inverse Higgs phenomenon \cite{ih}),
\be\label{ih}
\omega_Z=0 \Rightarrow \dot{q} = -\im \frac{\lambda}{1+\lambda\blambda} \und
\bar\omega_Z =0 \Rightarrow \dot{\bar q} = \im \frac{\blambda}{1+\lambda\blambda}
\ee
These constraints are purely kinematic ones. Thus, to realize this spontaneous breaking of $D=2{+}1$ Poincar\'{e}
symmetry we need two scalar fields, $q(t)$ and ${\bar q}(t)$.

Using the constraints \p{ih}, one may further simplify the Cartan forms \p{N4CF} to be
\be
\omega_P=\frac{1-\lambda\blambda}{1+\lambda\blambda}dt,\qquad \omega_T = \frac{d\lambda}{1-\lambda\blambda}, \qquad
{\bar\omega}_T = \frac{d\blambda}{1-\lambda\blambda}, \qquad
\omega_J = \im \frac{\lambda d\blambda-d\lambda \blambda}{1-\lambda\blambda}.
\label{omega}\ee

\subsection{Actions}
\begin{itemize}
\item
The simplest action, invariant under full $D=2{+}1$ Poincar\'{e} symmetry, is
\be\label{action1}
S_0=- m_0\int \omega_P = -m_0\int dt \sqrt{1- 4 {\dot q}\dot{\bar q}}\ .
\ee
It can be easily represented in Poincar\'{e}- and reparametrization-invariant form as
\be
S_0=-m_0\int d\tau\sqrt{\frac{dq^a}{d\tau} \frac{dq_a}{d\tau}}\equiv -m_0\int ds,\qquad q^0\equiv t,\quad \frac{q^1+\im q^2}{2}\equiv q\ ,
\ee
and for the summation we have used the Minkowski metric $g_{ab}=\textrm{diag}({+,-,-})$.
This is the action of a massive particle in $D=2{+}1$ spacetime.
\item

A less trivial action can be constructed as
\be\label{action2}
S_{\rm anyon}=-\frac{\alpha}{2} \int \omega_J =-\frac{\im\alpha}{2}\int dt \frac{  \dot\blambda \lambda -\blambda \dot\lambda}{1-\lambda\blambda}= \im\alpha\int dt \frac{{\ddot q} \dot{\bar q}-\dot{q} \ddot{\bar q}}{\sqrt{1- 4 {\dot q}\dot{\bar q}}\left(1+\sqrt{1- 4 {\dot q}\dot{\bar q}}\right)}.
\ee
In  reparametrization-invariant form it reads
\be
S_{\rm anyon}=\im\alpha\int\frac{(d^2q^1/d\tau^2) (dq^2/d\tau) -(d^2q^2/d\tau^2) (dq^1/d\tau)}{|dq^a/d\tau|(dq^0/d\tau+
|dq^a/d\tau|)}d\tau,\qquad q^0\equiv t\ .
\ee
It is seen that this defines the vector potential of a Dirac monopole in three-dimensional Minkowski space,
parameterized by the velocities $v^a\equiv dq^a/d\tau$.
Hence, we arrive at an action defining anyonic spin (see, e.g., \cite{anyon}).

\item
Finally, one may consider the action
\be\label{action3}
S_{\rm rigid} = \beta\int \frac{\omega_T \bar\omega_T}{\omega_P} = \beta\int dt \frac{1+\lambda\blambda}{(1-\lambda\blambda)^3}\dot\lambda\dot{\blambda} =
\beta\int dt \frac{\left( \dot{\bar q}\ddot{q}+\dot{q}\ddot{\bar q}\right)^2+(1-4{\dot q}\dot{\bar q})\ddot{q}\ddot{\bar q}}{(1-4 {\dot q}\dot{\bar q})^{5/2}}.
\ee
Representing this action in Poincar\'{e}- and reparametrization-invariant form, we get
\be
S_{\rm rigid}=\beta\int k^{2}_1({\dot q},\ddot{q})ds,
\ee
where
\be
k^{2}_1({\dot q},\ddot{q})\equiv \frac{({\ddot q}^a\dot{q}_a)^2-({\ddot q}^a{\ddot q}_a)(\dot{q}^b\dot{q}_b)}{({\dot q}^c{\dot q}_c)^{3}}
\qquad\textrm{with}\qquad {\dot q}^a=\frac{dq^a}{d\tau}, \quad{\ddot q}^a=\frac{d^2q^a}{(d\tau)^2}
\ee
is the square of the first extrinsic curvature (``rigidity'') of the worldline  in $\mathbb{R}^{1,2}$.
Note that systems  defined by the sum of \eqref{action1} and \eqref{action3} have been studied by various authors (see, e.g. \cite{ksquare}) .

\item
The most general action depending on $\lambda,\blambda$ and $\dot\lambda, \dot\blambda$ only
(i.e.~depending on up to second derivatives of $q$ and ${\bar q}$) has the form
\be\label{action_gen}
S_{\rm gen}=\int \frac{1-\lambda\blambda}{1+\lambda\blambda}{\cal F}\left[ \frac{(1+\lambda\blambda)^2 \dot\lambda \dot\blambda}{(1-\lambda \blambda)^4}\right] dt=\int   {\cal F}(k^{2}_1) ds
\ee
where $\cal F$ is an arbitrary function.
For the Hamiltonian analyses of such systems we refer to  \cite{tmp98}.
The most interesting case corresponds to the choice ${\cal F}(x)=c_0+c_1\sqrt{x}$,
i.e.~to a Lagrangian linear in the curvature, which has been studied extensively \cite{rigid}.
\end{itemize}
We remark that  $S_0$ and $S_{\rm rigid}$ as well as $S_{\rm gen}$ define Poincar\'{e}-invariant actions,
while $S_2$ is only weakly invariant under $D=2{+}1$ Poincar\'{e} transformations.

\subsection{Hamiltonian formulations}

In this subsection we shall consider the Hamiltonian formulation of the actions \eqref{action1}, \eqref{action2} and \eqref{action3} introduced in the previous subsection.

The Hamiltonian formulation of the action  \eqref{action1} is a textbook exercise.
In the static-gauge parametrization
it is defined by the symplectic structure $ dp \wedge dq + d{\bar p} \wedge d{\bar q} $ and by the Hamiltonian $p_0=\sqrt{m^2_0+p{\bar p}}$ and, obviously, it describes a $(2{+}1)$-dimensional scalar relativistic particle with mass $m_0$.

\subsubsection*{Majorana anyon}
Adding to \eqref{action1} the Wess-Zumino term \eqref{action2} provides the system with a non-zero spin but relaxes, at the classical level, the mass-shell condition.
So let us give the Hamiltonian formulation of $S=S_0+S_{anyon}$, in the static-gauge parametrization
\be
S=-m_0\int\omega_p-\frac{\alpha}{2}\int \omega_J=-m_0\int dt\sqrt{1- 4 {\dot q}\dot{\bar q}} + \im\alpha\int dt \frac{{\ddot q} \dot{\bar q}-\dot{q} \ddot{\bar q}}{\sqrt{1- 4 {\dot q}\dot{\bar q}}\left(1+\sqrt{1- 4 {\dot q}\dot{\bar q}}\right)}\ .
\ee
Taking into account the relations  \eqref{ih}
we rewrite its Lagrangian in a first-order form,
\be
{\tilde{\cal L}}=-m_0\frac{1-\lambda\blambda}{1+\lambda\blambda}-\frac{\im \alpha}{2} \frac{\lambda\dot{\blambda}-{\blambda}{\dot \lambda}}{1-\lambda\blambda} +p\left(\dot q+\frac{\im\lambda}{1+\lambda\bar \lambda}\right)+
{\bar p}\left(\dot{\bar q}-\frac{\im\blambda}{1+\lambda\blambda}\right).
\ee
This expression is of the form
${\tilde{\cal L}}={\cal A}_{(1)A}(x){\dot x}^A-{\cal H}(x)$, where $x^A=\{p, {\bar p}, \lambda,{\blambda}, q, {\bar q}\}$ are  independent variables,
\be
{\cal H}=p_0=\frac{\im{\bar p}{\blambda}-\im p \lambda
+m_0(1-\lambda\blambda)}{1+\lambda\blambda}
\label{H}\ee
is the Hamiltonian and
\be
{\cal A}_{(1)}=p d{ q} +{\bar p} d{\bar q}- \frac{\im\alpha}{2} \frac{\lambda d{\blambda}-{\blambda}{d \lambda}}{1-\lambda\blambda}
\label{A}\ee
is a one-form defining the symplectic structure
\be
\omega=d {\cal A}_{(1)}=dp\wedge dq + d{\bar p}\wedge d{\bar q}-\im\alpha \frac{d\lambda\wedge d\blambda}{(1-\lambda\blambda)^2}.
\label{ss0}\ee
This  symplectic structure defines Poisson brackets given by the non-zero relations
\be
\{p,q\}=1,\quad\{{\bar p},{\bar q}\}=1,\quad \{\lambda,\bar \lambda\}=
\frac{\im}{\alpha}(1-\lambda\blambda)^2.
\label{pb0}\ee

One can easily check that the generators of $so(1,2)$ are defined by
\be
J_0=2\im({\bar p}{\bar q}-p q )+\alpha \frac{1+\lambda\blambda}{1-\lambda\blambda}
\quad  J_+= {\bar p}+{ q}^2{ p} -\im\alpha\frac{\lambda}{1-\lambda\blambda}\;:\quad
\{J_\pm, J_0\}=2\im J_\pm,\quad \{J_+, J_-\}=\im J_0.
\ee
Together with
$p_0\equiv {\cal H}$, $p=(p_1+\im p_2)/2$, they form the $(2{+}1)$-dimensional Poincar\'{e} algebra.
The Casimirs of this algebra, $p_a p^a=:m^2$ and $p_aJ^a=:ms$, define
the spin $s$ and mass $m$ of the particle. Thus, we have the so-called Majorana condition
\be
ms=m_0\alpha=\textrm{const},
\ee
i.e.~we deal with a reducible representation of the Poincar\'{e} group.
This $(2{+}1)$-dimensional system has been studied in detail in~\cite{plyushchay},
where it was called a ``Majorana anyon''.
We remark that the Lagrangian of~\cite{plyushchay} featured a linear dependence on the second extrinsic curvature (torsion) $\kappa_2$ and thus included third-order time derivatives as well.
A Majorana anyon can also be described by a simple second-order action on null curves~\cite{ramos}.

\subsubsection*{Rigid particle}

Let us give a Hamiltonian formulation for the action containing a rigidity term quadratic in the first extrinsic curvature,
\be
S=S_0+S_{\rm anyon}+S_{\rm rigid}=\int{\cal L}dt ,\qquad
{\cal L}=-m_0\omega_P-\frac{\alpha}{2}\Omega_J+\beta\frac{\omega_T{\bar\omega}_T}{\omega_P}.
\ee
Its Poincar\'{e}-covariant formulation (in the absence of an anyonic term, i.e.~for $\alpha=0$) is well-known and has been considered by many authors \cite{ksquare,tmp98}.
Here, we restrict ourselves to the Hamiltonian formulation in the static gauge.
In complete analogy with the previous case, we replace the Lagrangian by an equivalent first-order one,
\be
{\tilde{\cal L}}=-m_0\frac{1-\lambda\blambda}{1+\lambda\blambda}-\frac{\alpha}{2} \frac{\im(\lambda\dot{\blambda}-{\blambda}{\dot \lambda})}{1-\lambda\blambda} +
\Pi{\dot\lambda}+{\bar\Pi}{\dot\blambda}-\frac{1}{\beta}
 \frac{(1-\lambda\blambda)^3}{1+\lambda\blambda}\Pi{\bar\Pi}
+p\left(\dot q+\frac{\im\lambda}{1+\lambda\blambda}\right)+
{\bar p}\left(\dot{\bar q}-\frac{\im\blambda}{1+\lambda\blambda}\right)\ .
\ee
Hence, the system is described by the Hamiltonian
\be
{\cal H}_{rigid}=p_0=\frac{1-\lambda\blambda}{1+\lambda\blambda}\left(\frac{1}{\beta} (1-\lambda\blambda )^2\Pi{\bar\Pi}-
\frac{\im(p\lambda-{\bar p}\blambda)}{1-\lambda\blambda}+m_0
\right)
\label{po}\ee
and by the symplectic one-form
\be
{\cal A}_{(1)}=p d{ q} +{\bar p} d{\bar q}+\Pi d{\lambda} +{\bar\Pi} d{\blambda} - \frac{\im\alpha}{2} \frac{\lambda d{\blambda}-{\blambda}{d \lambda}}{1-\lambda\blambda}.
\ee
The latter yields the symplectic structure
\be
\omega=d{\cal A}_{(1)}=dp\wedge dq +d{\bar p} \wedge d{\bar q}+
d\Pi\wedge d\lambda +d{\bar\Pi}\wedge d\blambda - {\im \alpha}\frac{d\lambda\wedge d\blambda}{(1-\lambda\blambda )^2},
\ee
and the corresponding non-zero Poisson brackets
\be
\{p,q\}=1,\quad\{{\bar p},{\bar q}\}=1,\quad  \{\Pi, \lambda\}=1, \quad
\{{\bar\Pi},\blambda\}=1,\quad
\{\Pi,{\bar\Pi}\}=-{\im}{\alpha}(1-\lambda\blambda)^2.
\label{pb00}\ee
The Lorentz generators read
\be
J_0=2\im({\bar p}{\bar q}-p q )+2\im({\bar\Pi}{\blambda} -\Pi\lambda)
+{\alpha} \frac{1+\lambda\blambda}{1-\lambda\blambda}
\und  J_+= {\bar p}+{ q}^2{ p}+
{\bar\Pi}+{\lambda}^2{\Pi}-\im\alpha\frac{\lambda}{1-\lambda\bar \lambda},
\ee
while the translation generators are given, as before, by $\{p_0={\cal H}_{rigid}, p,{\bar p}\}$.
It is easy to check that neither spin nor mass are fixed in this model.

\setcounter{equation}{0}
\section{Supersymmetric generalization}
In this section we turn to ${\cal N}{=}4$ supersymmetric extensions of the actions given above.
Two of the four supercharges are assumed to be spontaneously broken,
leaving us with ${\cal N}{=}2$ unbroken supersymmetry.

\subsection{Coset approach: kinematics}
We begin with the ${\cal N}{=}2, D{=}2{+}1$ super-Poincar\'e algebra, which in $d=1$ notation appears as ${\cal N}{=}4, d{=}1$ super-Poincar\'{e}
algebra with two central charges. The basic  (anti)commutation relations extend the previous relations \p{N4d1} by
\bea\label{N4d1susy}
&& \left\{ Q, \bQ\right\} = 2 P, \; \left\{ S, \bS\right\} = 2 P, \quad
\left\{ Q, S\right\} = 2 Z, \; \left\{ \bQ, \bS\right\} = 2 \bZ, \nn \\
&& \begin{array}{l}
\left[ J, Q\right] =\frac{1}{2} Q,\; \left[ J, \bQ\right] =-\frac{1}{2} \bQ, \\
\left[J,S\right] =\frac{1}{2} S, \;\left[ J, \bS\right] =-\frac{1}{2} \bS,
\end{array}
\qquad
\begin{array}{l}
\left[ T, \bQ\right] = - S, \\
\left[ T ,\bS\right] =- Q ,
\end{array}
\qquad
\begin{array}{l}
\left[ \bT, Q\right] = \bS, \\
\left[ \bT , S\right] = \bQ .
\end{array}
\eea
Here, $Q,\bQ$ and $S,\bS$ are the generators of the unbroken and spontaneously broken supersymmetries, respectively.
$P$ is the generator of translation,  $Z,\bZ$ are the central charge generators, while $T,\bT,J$ are the generators of
the $D=2{+}1$ Lorentz group, as before.

In the coset approach \cite{coset1,coset2}, the breakdown of $S$ supersymmetry and $Z,\bZ$ translations is reflected in the structure of the coset element
\bea\label{N4d1coset}
g= e^{\im t P}\; e^{\theta Q +\bar\theta \bQ}\; e^{\mpsi S +\bar\mpsi \bS}\; e^{\im\left( \mq Z + {\bar \mq} \bZ\right)} \; e^{\im\left( \mLambda T + \mbLambda \bT\right)}.
\eea
The ${\cal N}=2$ superfields $\mq( t, \theta,\bar\theta), \mpsi( t, \theta,\bar\theta)$ and $\mLambda( t, \theta,\bar\theta)$ are Goldstone superfields accompanying the ${\cal N}{=}2, D{=}2{+}1$ super-Poincar\'{e} to ${\cal N}{=}2, d{=}1$ super-Poincar\'{e} breaking.

The transformation properties of the coordinates and superfields are induced by the left multiplications of the coset element  \p{N4d1coset},
\be
g_0\; g = g' \; h, \qquad h\sim e^{f J}.
\ee
The most important transformations read
\begin{itemize}
\item Unbroken SUSY\;$ \left( g_0 = e^{\epsilon Q +\bar\epsilon \bQ}\right) :\qquad\delta\theta =\epsilon,\quad \delta t =\im\left( \epsilon \bar\theta+\bar\epsilon\theta\right); $
 \item Broken SUSY \;$ \left( g_0 = e^{\varepsilon S +\bar\varepsilon \bS}\right) :\qquad
 \delta t = \im \left( \varepsilon \bar\mpsi+\bar\varepsilon \mpsi\right),\quad \delta \mpsi =\varepsilon,\quad \delta \mq = 2 \im\; \varepsilon\theta,$
 \item Automorphism group\;$ \left( g_0 = e^{\im\left( \alpha T +\bar\alpha \bT\right)}\right):\qquad
  \left\{
 \begin{array}{l}
 \delta t = - 2 \im\left( \alpha {\bar \mq}-{\bar\alpha}\mq\right) +2 \alpha \bar\theta\bar\mpsi -2 \bar\alpha \theta\mpsi, \quad
 \delta\theta =-\im \alpha\bar\mpsi,\\
 \delta \mq =  \alpha\left(-\im t-\theta\bar\theta+\mpsi\mbpsi\right),\quad \delta\mpsi = -\im \alpha\bar\theta,\quad
 \delta\mlambda =\alpha-\bar\alpha\mlambda{}^2,
 \end{array}
 \right. $
 \end{itemize}
where, as in \p{N4def2} before,
\be
\mlambda = \frac{ \tanh (\sqrt{\mLambda \mbLambda})}{\sqrt{\mLambda \mbLambda}}\mLambda\ .
\ee

The left invariant Cartan forms read

\bea\label{N4d1CF}
& \omega_P = \frac{1}{1-\mlambda\mblambda }
\left[ \left(1+\mlambda\mblambda\right) \triangle t + 2 \im \left( \mlambda \triangle{\bar \mq}-\mblambda \triangle \mq \right)\right],
\qquad \omega_Z = \frac{1}{1-\mlambda\mblambda}\left[\triangle \mq -\mlambda^2 \triangle {\bar \mq} +\im \mlambda \triangle t\right],\;
&\nn \\
& \omega_T = \frac{d\mlambda}{1-\mlambda\mblambda},\qquad
 \omega_J =\im \frac{\mlambda d\mblambda - d\mlambda \mblambda}{1-\mlambda \mblambda}.&\\
 &\omega_Q = \frac{1}{\sqrt{1-\mlambda\mblambda}}\left[ d \theta +\im \mlambda d\mbpsi\right], \qquad
 \omega_S = \frac{1}{\sqrt{1-\mlambda\mblambda}}\left[ d \mpsi +\im \mlambda d\bar\theta\right]. &\nn
\eea
Here,
\be\label{N4d1susydif}
\triangle t  = dt - \im \left( \theta d \bar\theta +\bar\theta d \theta + \mpsi d \mbpsi +\mbpsi d \mpsi\right) \und
\triangle \mq = d \mq - 2 \im \mpsi d\theta .
\ee
Having at hands the Cartan forms, one may construct ``semi-covariant'' derivatives (covariant with respect to $P,J$, broken and
unbroken supersymmetries, only) via
\be\label{cD}
\triangle t \nabla_t + d\theta \nabla_\theta +d\bar\theta \bnabla_{\theta} = dt \frac{\partial}{\partial t}+
d\theta \frac{\partial}{\partial\theta} + d\bar\theta \frac{\partial}{\partial\bar\theta}.
\ee
Explicitly, they read
\be\label{N4d1scd}
\nabla_t =E^{-1} \partial_t, \quad \nablath = D -\im \left( \mbpsi D \mpsi + \mpsi D\mbpsi\right) \nabla_t,\quad
\bnablath=\bD -\im\left( \mbpsi \bD \mpsi + \mpsi \bD\mbpsi\right) \nabla_t,
\ee
where
\be E= 1 + \im\left( \dot\mpsi \mbpsi +  \dot\mbpsi \mpsi\right), \qquad
 D= \frac{\partial}{\partial \theta } -\im \bar\theta \partial_t,\quad\bD= \frac{\partial}{\partial \bar\theta } -\im \theta \partial_t\; :  \quad  \left\{D, \bD \right\}=-2\im \partial_t.
 \ee
These derivatives obey the following algebra,
\bea\label{N4d1der}
&& \left\{\nablath, \bnablath\right\} =-2 \im\left( 1+  \nablath\mpsi\bnablath\mbpsi +
\bnablath\mpsi\nablath \mbpsi\right)  \nabla_t , \nn \\
&& \left\{ \nablath, \nablath\right\} = -4 \im \nablath\mbpsi \nablath\mpsi\nabla_t, \quad
\left\{ \bnablath, \bnablath\right\} = -4 \im \bnablath\mbpsi \bnablath\mpsi\nabla_t, \\
&& \left[ \nabla_t, \nablath\right] = - 2\im \left( \nablath \mbpsi \nabla_t \mpsi +\nablath \mpsi \nabla_t\mbpsi \right) \nabla_t, \quad
\left[ \nabla_t, \bnablath\right] = - 2\im \left( \bnablath \mbpsi \nabla_t \mpsi +\bnablath\mpsi \nabla_t\mbpsi \right) \nabla_t .\nn
\eea
Finally, imposing the same constraints \p{ih} as in the bosonic case, one may reduce the number of independent Goldstone superfields,
\bea\label{N4d1constr}
\begin{array}{l}
\omega_Z=0 \quad\Rightarrow \quad  \nabla_t {\mq} = -\im \frac{\mlambda}{1+\mlambda\mblambda}, \; \nablath \mq + 2\im \mpsi =0, \bnablath \mq =0, \\
\bar\omega_Z =0 \quad\Rightarrow \quad \nabla_t {\bar \mq} = \im \frac{\mblambda}{1+\mlambda\mblambda}, \;  \bnablath \bar\mq + 2\im \mbpsi =0, \nablath \bar\mq =0.
\end{array}
\eea
These constraints impose covariant chirality conditions on the superfields $\mq$ and $\bar\mq$ and, in addition,
they express the Goldstone superfields $\mpsi,\mbpsi,\mlambda, \mblambda$ as the derivatives of the $\mq$ and $\bar\mq$,
thereby realizing the inverse Higgs effect \cite{ih}. Thus, we have in the system only one, covariantly chiral, ${\cal N}=2$ complex bosonic superfield $\mq(t,\theta,\bar\theta)$.

The constraints \p{N4d1constr} imply some further restrictions. For example, if we act by $\nablath$ on the constraint $\nablath \mq + 2\im \mpsi =0$, we will get
\be\label{N4d1addcons1}
\nabla_\theta^2 \mq + 2 \im \nabla_\theta \mpsi= 0 \Rightarrow 2\im \nabla_\theta \mpsi \left( 1 - \nabla_\theta \mbpsi \nabla_t \mq\right)=0.
\ee
Thus, we have to conclude that
\be\label{conn}
 \nablath \mpsi =0.
\ee
Moreover, on the constraint surface given by \p{N4d1constr} and \p{conn} the algebra of covariant derivatives slightly simplifies:
\be\label{N4d1constr2}
 \nablath^2 = \bnablath^2=0, \; \left\{\nablath, \bnablath \right\} = -2 \im \left( 1 +\bnablath\mpsi \nablath\mbpsi \right)  \nabla_t,  \; \left[ \nabla_t, \nablath \right] = - 2\im  \nablath \mbpsi \nabla_t \mpsi  \nabla_t, .
\ee

\subsection{Component transformation laws}
As we are going to define component actions, we need transformation laws for the components. Let us firstly denote the components of superfields in the following way,
\be\label{N4d1comps1}
\mq|_{\theta=0} = q, \; \bar \mq|_{\theta=0} = \bq,\qquad
\mpsi |_{\theta=0} = \psi, \; \mbpsi|_{\theta=0} = \bpsi,\qquad
\mlambda|_{\theta=0} = \lambda, \; \mblambda|_{\theta=0} = \blambda.
\ee
It appears to be convenient to introduce also the quantity
\be
\cE =E|_{\theta=0}= 1+ \im \left( \dot\psi \bpsi + \dot\bpsi \psi  \right)
\ee
and to define a new time derivative,
\be
\cD_t = \cE^{-1}\partial_t .
\ee

We list the active transformation laws (at fixed $t$) for these components under the broken and unbroken supersymmetries.

Broken supersymmetry:
\be\label{N4d1cmpbs}
\delta^\star_S q = - \im \left(\varepsilon \bpsi + \bar\varepsilon \psi  \right)\dot q, \quad
\delta^\star_S \psi =\varepsilon - \im \left(\varepsilon \bpsi + \bar\varepsilon \psi  \right)\dot \psi, \qquad
\delta^\star_S \cE = -\im \partial_t\left[ \cE \left(\varepsilon \bpsi + \bar\varepsilon \psi  \right)  \right].
\ee

Unbroken supersymmetry $\delta^\star_Q f|_{\theta=0}=\left( \epsilon D f + \bar\epsilon \bar D f\right)|_{\theta=0}$:
\bea\label{N4d1cmpus}
&&\delta^\star_Q q = -2\im \epsilon \psi + \left(\bar\epsilon \bpsi \lambda - \epsilon \psi \blambda \right)\dot q,\quad
\delta^\star_Q \psi =-\im \bar\epsilon \lambda + \left(\bar\epsilon \bpsi \lambda - \epsilon \psi \blambda \right)\dot \psi,\nn \\
&& \delta^\star_Q \cE =  \partial_t\left[ \cE \left(\bar\epsilon \bpsi \lambda - \epsilon \psi \blambda \right)  \right] + 2\left( \epsilon \dot\psi \blambda - \bar\epsilon \dot\bpsi \lambda \right).
\eea

Finally, we stress that the relations between the components $\lambda$ and $q$ are given by the following expressions,
\be\label{add1}
\cD_t q = -\im \frac{\lambda}{1+\lambda\blambda}  ,\qquad
 \Leftrightarrow \qquad \lambda = 2\im \frac{\cD_t q}{1+ \sqrt{1-4\cD_t q \cD_t {\bar q}}} .
\ee
\subsection{Actions}
We are ready to construct the supersymmetric generalization of the actions \p{action1}, \p{action2} and \p{action3}. As they have different dimensions, these actions must be invariant separately.

\subsubsection*{Superparticle}
It is easy to check that the evident ansatz
\be\label{susy1}
\int dt \cE F_1(\lambda\blambda)
\ee
for the supersymmetric extension of the particle action \p{action1}
is perfectly invariant with respect to the broken supersymmetry \p{N4d1cmpbs}, because
\bea\label{actions1a}
\delta^\star_S \left( \cE F_1(\lambda\blambda) \right) = -\im \partial_t\left[ \cE \left(\varepsilon \bpsi + \bar\varepsilon \psi  \right)  \right] F_1 - \im \cE \left(\varepsilon \bpsi + \bar\varepsilon \psi  \right) \left( \lambda \dot{\blambda} + \dot \lambda \blambda \right) F'_1 = -\im \partial_t\left[ \cE F_1 \left(\varepsilon \bpsi + \bar\varepsilon \psi  \right)  \right].
\eea
To determine the function $F_1(\lambda\blambda)$, we impose invariance under the unbroken supersymmetry \p{N4d1cmpus}.
The corresponding variation of $ \cE F_1(\lambda\blambda)$ computes to
\be\label{actions2}
\delta^\star (\cE F_1)= -\partial_t \left[ \cE\left( \epsilon \psi \blambda - \bar\epsilon \bpsi \lambda  \right) F_1 \right] + 2\left(\epsilon \dot\psi \blambda - \bar\epsilon \dot\bpsi \lambda \right)\left( F_1 + (1+\lambda \blambda)F_1^\prime \right).
\ee
The first term of this variation is a total time derivative, while the second one is not. It is absent, however, for $F_1 \sim \left( 1+\lambda\blambda \right)^{-1}$.
So, choosing $F_1=-\frac{2m_0}{1+\lambda\blambda}$, our ansatz \eqref{susy1} produces a supersymmetic action.

Then, we directly get the invariant supersymmetric extension of the action \eqref{action1} as
\bea\label{actions1}
{\cal S}_0 = m_0\int dt -2m_0\int\frac{ \cE dt}{1+\lambda\blambda}  =
-m_0\int dt \left[ \cE  \sqrt{1-4 \cD_t q \cD_t \bq}  +\cE -1 \right].
\label{0s}\eea
This is just the action of the ${\cal N}{=}2, D{=}2{+}1$ superparticle in the form considered in \cite{BKKS1}.
Having in mind the relations \eqref{add1}, one may rewrite the Lagrangian in the form
\be
{\cal L}_0=-m_0\sqrt{\cE ^2-4{\dot q}\dot{\bar q}}- m_0(\cE-1).
\ee

Let us give the Hamiltonian formulation of this system.
The momenta $p$, $\pi$  conjugate to $q$, $\psi$ read
\be
p=\frac{2m_0{\dot{\bar q}}}{\sqrt{\cE ^2-4{\dot q}\dot{\bar q}}}
\und \pi=\im m_0\left( \frac{\cE}{\sqrt{\cE ^2-4{\dot q}\dot{\bar q}}}+1\right)\bpsi,
\ee
from where we immediately get the Hamiltonian
\be
{\cal H}_0=\sqrt{m^2_0+p\bar p}
\ee
and fermionic constraints
\be
\pi=\im( m_0 +\sqrt{m^2_0+p\bar p}){\bar\psi}
\und  {\bar\pi}=-\im( m_0 +\sqrt{m^2_0+p\bar p}){\psi}.
\ee
Substituting these expressions   into the symplectic one-form $
{\cal A}_1=p{d q}+{\bar p}{d{\bar q}}+ \pi{d\psi}-{\bar\pi}{d{\bar \psi}}$,
it reduces to
\be
{\cal A}_{\rm red}=pd q+{\bar p}{d{\bar q}}+\im \left( m_0 +\sqrt{m^2_0+p\bar p}\right)\left(\psi d{\bar\psi} + {\bar\psi}d\psi\right)\ .
\label{ared}\ee
{}From the  symplectic structure $d{\cal A}_{\rm red}$, we read off the Poisson brackets
defined by the non-zero relations
\be
\left\{ p,q\right\}=1,\quad \left\{\psi,\bpsi\right\}=\frac{\im}{2\left(m_0+\cH_0\right)},\quad
\left\{\psi, q\right\}=-\frac{\psi{\bar p}}{4\left(m_0+\cH_0\right)\cH_0},\quad
\left\{\psi, {\bar q}\right\}=-\frac{\psi p}{4\left(m_0+\cH_0\right)\cH_0}.
\ee
The transformation properties \p{N4d1cmpbs}, \p{N4d1cmpus} then tell us the supercharges
\be\label{supercharges}
Q= 2 p \psi, \quad S= 2\left(m_0+\cH_0\right) \bpsi.
\ee
Indeed, these forms of $Q$ and $S$ produce the proper shifts of $q$ and $\psi$,
respectively,
\be
\delta^\star_Q q=-\im \epsilon \left\{Q,q\right\} \sim -2 \im \epsilon \psi+\ldots \und
\delta^\star_S \psi =-\im \varepsilon \left\{S,\psi\right\}=\varepsilon.
\ee
It is matter of straightforward calculations to check that the remaining terms in \p{N4d1cmpbs} and \p{N4d1cmpus} are also reproduced.

The supercharges \p{supercharges} form centrally extended ${\cal N}{=}4, d{=1}$ super-Poincar\'{e} algebra,
\be
\{Q,\bQ\}=2\im \left(\cH_0-m_0\right),\qquad \{S,\bS\}=2\im \left(\cH_0+m_0\right),\qquad \{Q, S\}=2\im p.
\ee
{}From \eqref{ared} we can readily deduce the canonical coordinates $p$ and
\be
\chi=\sqrt{m_0 +\cH_0}\;\psi,\quad {\tilde q}=q-\im\frac{{\bar p}}{\sqrt{m_0 +\cH_0}}(\psi\bar\psi):\qquad
\{p,{\tilde q}\}=1,\qquad\{\chi,\bar\chi\}=-\frac{\im}{2}.
\ee
In these coordinates,   the supercharges read
\be
Q=2\frac{p\chi}{{\sqrt{m_0 +\cH_0}}} \und
S=2{\sqrt{m_0 +\cH_0}}\;{\bar\chi}\ .
\ee

Finally, we note that the action \p{actions1} can be written in terms of superfields as
\be
{\cal S}_0=2m_0\int dt d\theta d{\bar\theta}\frac{\mpsi{\mbpsi}}{1+\mlambda\mblambda}.
\ee

\subsubsection*{Supersymmetric anyon}
The supersymmetrization of the anyonic  action \p{action2}  is more involved. The most general ansatz with the proper bosonic limit reads\footnote{
The second term in \p{actions3} is of the proper dimension but disappears in the bosonic limit.}
\be\label{actions3}
{\cal S}_{\rm anyon}=\frac{\im\alpha}{2} \int dt\; \cE \frac{ \cD_t \lambda \blambda -\lambda \cD_t\blambda}{1-\lambda\blambda}+
\int dt\; \cE \;F_2(\lambda\blambda) \cD_t \psi \cD_t \bpsi.
\ee
This action is invariant with respect to the broken supersymmetry \p{N4d1cmpbs} because
\be\label{bsusy11}
\delta^\star_S \left[ \im \; \cE\; \frac{ \cD_t \lambda \blambda -\lambda \cD_t\blambda}{1-\lambda\blambda}\right]
=\partial_t\left[\left(\varepsilon \bpsi + \bar\varepsilon \psi  \right)\;\cE\; \frac{  \cD_t \lambda \blambda - \cD_t{\blambda} \lambda  }{1-\lambda\blambda}   \right]
\ee
and
\be\label{bsusy12}
\delta^\star_S \left[ \cE\; \cD_t\psi \cD_t\bpsi F_2  \right] = -\im \partial_t \left[   \left(\varepsilon \bpsi + \bar\varepsilon \psi  \right) \cE\; \cD_t\psi \cD_t\bpsi F_2  \right].
\ee
A straightforward calculation shows that invariance under unbroken supersymmetry fixes $F_2$ to
\be\label{actions9}
F_2 = -2\alpha\frac{1+\lambda\blambda}{(1-\lambda\blambda)^2},
\ee
and the full supersymmetric anyonic action acquires the form
\bea\label{susyan}
{\cal S}_{\rm anyon} &=& \frac{\im\alpha}{2}\int dt \;\cE\;\frac{\blambda \cD_t\lambda - \lambda\cD_t{\blambda}}{1-\lambda\blambda} - 2\alpha\int dt \;\cE\; \frac{1+\lambda\blambda}{(1-\lambda\blambda)^2} \cD_t\psi \cD_t\bpsi  \nn \\
&=& \im\alpha \int dt\; \cE\; \frac{\cD_t (\cD_t q) \cD_t \bq -\cD_t (\cD_t \bq) \cD_t q }{\sqrt{1-4\cD_t q \cD_t \bq} \left( 1+\sqrt{1-4\cD_t q \cD_t \bq} \right)} -\alpha\int dt\; \cE\;\frac{1+ \sqrt{1-4\cD_t q \cD_t \bq}}{1-4\cD_t q \cD_t \bq} \cD_t\psi \cD_t\bpsi .
\label{2s}\eea

Two notes are in order:
\begin{itemize}
\item The forms $\omega_S$ and $\bar\omega_S$ can be evaluated on the superfield constraints \p{N4d1constr2}, which removes the $d\theta$ and $d\bar\theta$ projections.
We find that the $\dot\psi \dot\bpsi$ term can be represented as
    \bea\label{actions10}
    -2\int \frac{\omega_S | \cdot \bar\omega_S |}{\omega_P|}.
    \eea
\item The superfield expression for the action \p{susyan} takes the simple form
\be
{\cal S}_{\rm anyon}=\frac{\im \alpha}{2}\int dt d\theta d{\bar\theta}\frac{{\dot\mpsi \mbpsi+{\dot{\mbpsi}\mpsi}}}{1-\mlambda\mblambda}.
\ee
\end{itemize}

We are ready to give a Hamiltonian formulation of the supersymmetric extension of the  anyonic system.
It is defined as the sum of the particular actions \eqref{0s} and \eqref{2s},
${\cal S}={\cal S}_0+{\cal S}_{\rm anyon}$.
Introducing fermionic  momenta $\eta$ and $\bar\eta$ conjugate to
the Grassmann variables $\psi$ and $\bar\psi$,
the first-order Lagrangian reads
\be
{\tilde{\cal L}}=m_0-\frac{2m_0{\cal E}}{1+\lambda\blambda}-
\frac{\im\alpha}{2}\frac{{\lambda}{\dot\blambda}-{\blambda}{\dot\lambda}}{1-\lambda\blambda}
+\eta{\dot\psi}-{\bar\eta}\dot{\bar\psi}
-\frac{1}{2\alpha}\frac{(1-\lambda\blambda)^2{\eta\bar\eta}}{1+\lambda\blambda}
+p\left({\dot q}
+\frac{\im{\cal E}\lambda}{1+\lambda\blambda}\right)+{\bar p}\left({\dot{\bar q}}
-\frac{\im{\cal E}\blambda}{1+\lambda\blambda}\right).
\ee
Hence, the Hamiltonian is given by the expression
\be
{\cal H}_{SUSY}={\cal H} +\frac{1}{2\alpha}\frac{(1-\lambda\blambda )^2\eta{\bar\eta}}{1+\lambda\blambda},
\ee
where ${\cal H}$ is defined in \eqref{H} as
\be
{\cal H}=
\im\frac{{\bar p}\blambda- p\lambda}{1-\lambda\blambda}+m_0\frac{1-\lambda\blambda}{1+\lambda\blambda} .
\ee

The symplectic structure follows from the one-form
\be
{\cal A}_1=pd q+{\bar p}{d{\bar q}} -\frac{\im\alpha}{2}\frac{d\blambda \lambda-d\lambda \blambda}{1-\lambda\blambda}+
\pi d\psi-{\bar\pi} d{\bar\psi},
\ee
where
\be
{\pi}=\eta-\im({\cal H}+m_0)\bar\psi \und  {\bar\pi}=\bar\eta+\im({\cal H}+m_0)\psi\ .
\ee
Therefore, the Poisson brackets are defined by the relations
\be
\{p,q\}=1,\quad \{\lambda,\blambda\}=\frac{\im}{\alpha}\left(1-\lambda\blambda\right)^2,\quad\{\pi,{\psi}\}=1,\quad
 \{{\bar\pi},{\bar\psi}\}=-1\ .
\ee
In these terms the Hamiltonian and supercharges read
\bea
&&{\cal H}_{SUSY}={\cal H}+\frac{1}{2\alpha}\frac{(1-\lambda\blambda)^2}{1+\lambda\blambda}
 \left(\pi+\im({\cal H}+m_0)\bar\psi\right)\left({\bar\pi}-\im({\cal H}+m_0)\psi\right)\ ,\\
&&Q=2 p\psi+\blambda \left({\bar\pi}-\im({\cal H}_{SUSY}+m_0)\psi\right),\quad S=\im\pi +{\bar\psi} ({\cal H}_{SUSY}+m_0).
\eea
They form the superalgebra
\be
\{Q,{\bar Q}\}=2\im({\cal H}_{SUSY}-m_0) ,\quad \{S,{\bar S}\}=2\im ({\cal H}_{SUSY}+m_0),\qquad \{Q,{S}\}=2\im p.
\ee

\subsubsection*{Rigid superparticle}
The supersymmetric extension of the bosonic term
\be
\int dt \frac{1+\lambda\blambda}{(1-\lambda\blambda)^3} \dot\lambda\dot\blambda
=:\int dt\ G_1(\lambda\blambda)\ \dot\lambda\dot\blambda
\ee
from \p{action3} is a more complicated task, due to the existence of two further expressions of the proper dimension, which however vanish in the bosonic limit, namely
\be\label{actionrp1}
\im G_2 (\lambda\blambda) \left( \ddot\psi \dot\bpsi + \ddot \bpsi \dot\psi \right) \und
\im G_3(\lambda\blambda) \left( \dot \lambda \blambda - \dot{\blambda} \lambda  \right)\dot\psi \dot\bpsi.
\ee
All three terms can be immediately promoted to be invariant under the broken supersymmetry, giving
\be\label{actionrp2}
{\cal S}_{\rm rigid} = \int \cE dt \left[ G_1 \cD_t\lambda\cD_t\blambda +\im G_2 \left( \cD_t^2 \psi \cD_t \bpsi + \cD_t^2 \bpsi \cD_t\psi \right) + \im G_3 \left( \cD_t \lambda  \blambda - \cD_t{\lambda}  \lambda  \right)   \cD_t \psi \cD_t\bpsi\right],
\ee
where we temporarily unfix the function $G_1$. We expect the three functions $G_1$, $G_2$ and $G_3$ to be constrained by invariance under unbroken supersymmetry.

After quite lengthy calculations, we find that our action
\be\label{Srig}
 {\cal S}_{\rm rigid}=\int dt \left[  G_1 \cE^{-1} \dot \lambda \dot{\blambda} + \im G_2 \cE^{-2} \left( \ddot\psi \dot\bpsi + \ddot \bpsi \dot\psi \right) + \im G_3 \cE^{-2} \left( \dot \lambda  \blambda - \dot{\blambda}  \lambda  \right) \dot\psi \dot\bpsi\right]
\ee
is invariant under unbroken supersymmetry if the equations
\be\label{actionrp7}
-G_3 + G_2^\prime +2 G_1 =0,\quad
G_3 + G_2^\prime+2 (1+\lambda\blambda) G_{1}^\prime =0,\quad
G_2 + (1+\lambda\blambda) G_1=0
\ee
hold, where the prime denotes a derivative with respect to the single argument $\lambda\blambda$ of these functions.
These equations are not independent, because the sum of first two reduces to the derivative of the third.
The solution of this system reads
\be\label{rp1}
G_2 = - (1+\lambda\blambda)G_1 \und G_3 = G_1 - (1+\lambda\blambda)G_1^\prime.
\ee
Thus, invariance with respect to both ${\cal N}{=}2$ supersymmetries determines the action up to one arbitrary function $G_1(\lambda\blambda)$. The prescribed bosonic limit fixes this function to
\be
G_1=\frac{1+\lambda\blambda}{(1-\lambda\blambda)^3},
\ee
and thus the complete ${\cal N}{=}4$ supersymmetric generalization of the rigid-particle action has the form
\be\label{actionrp8}
{\cal S}_{\rm rigid} =\int dt \left[ \frac{1+\lambda\blambda}{(1-\lambda\blambda)^3} \cE^{-1}\dot \lambda \dot{\blambda} - \im\frac{(1+\lambda\blambda)^2}{(1-\lambda\blambda)^3}\cE^{-2}\left( \ddot\psi \dot\bpsi + \ddot \bpsi \dot\psi \right) - 3\im \frac{(1+\lambda\blambda)^3}{(1-\lambda\blambda)^4}\left( \dot\lambda \blambda - \dot{\blambda} \lambda \right)\cE^{-2} \dot\psi \dot\bpsi\right].
\ee
In superfield language this action can be written in the much more compact form
\be
{\cal S}_{\rm rigid} =\int dt d\theta d{\bar\theta}\frac{1+\mlambda\mblambda}{(1-\mlambda\mblambda)^3}{\dot\mpsi}{\dot{\bar\mpsi}}.
\ee
The Hamiltonian formulation of the supersymmetric rigid particle will be considered elsewhere.

\newpage

\section{Discussion and outlook}
We have applied the coset approach to the construction of component actions describing a superparticle in $D{=}2{+}1$ spacetime, with ${\cal N}{=}4$ supersymmetry partially broken to ${\cal N{}}=2$, and with the bosonic action containing higher time derivatives, in the forms of an anyonic term and the square of the first extrinsic curvature. We presented the supercharges for the unbroken and broken supersymmetries as well as the Hamiltonian for the supersymmetric anyon and provided the superspace actions for all cases.

Our main goal was to find out whether it is possible to apply the approach, previously developed for the
construction of supersymmetric actions with a minimal number of time derivatives \cite{BKS1,BKKS1},
also to systems with higher time derivatives in the bosonic sector.
We are aware that the simple ${\cal N}{=}4 \rightarrow {\cal N}{=}2$ pattern of supersymmetry breaking
drastically simplifies the analysis (for example, by the absence of auxiliary components). Clearly, the analysis of more involved systems with higher supersymmetries or higher target-space dimensions is desired.
Using the fermions of the nonlinear realization as the physical fermionic components renders
the constructed actions quite compact and involves only geometric objects such as the einbein and
covariant derivatives of the bosonic ``matter'' fields and the fermions.

An interesting further question is whether also $p$-brane actions (with $p\geq 1$) containing higher derivatives can be supersymmetrized in a similar way.
Such a generalization is not obvious, however, due to presence of auxiliary fields, which have to be excluded by their, a priori unknown, equations of motion.

\vspace{0.4cm}
\noindent{\bf Acknowledgments.}
S.K. and A.N. are grateful to Leibniz Universit\"at Hannover for warm hospitality.
This work was partially supported by RFBR grants~12-02-00517-a, 13-02-91330-NNIO-a and 13-02-90602 Apm-a,
by  grants of Armenian State Committee of Science 13RF-018,  13-1C114,
and by the Volkswagen Foundation under the grant I/84 496.


\begin{thebibliography}{99}

\bibitem{coset1} S.R.~Coleman, J.~Wess, B.~Zumino,\\
{\it Structure of phenomenological lagrangians. 1},
Phys.\ Rev.\ {\bf 177} (1969) 2239;\\
{\it Structure of phenomenological lagrangians. 2},
Phys.\ Rev.\ {\bf 177} (1969) 2247.
\bibitem{coset2} D.V.~Volkov,
{\it Phenomenological lagrangians},
Sov.\ J.\ Part.\ Nucl.\ {\bf 4} (1973) 3.

V.I.~Ogievetsky,
{\it Nonlinear realizations of internal and space-time symmetries},\\
in: Proceedings of the Xth Winter School of Theoretical Physics in Karpacz, Vol.1, p.117, 1974.
\bibitem{ih} E.A.~Ivanov, V.I.~Ogievetsky,
{\it The inverse Higgs phenomenon in nonlinear realizations},\\
Teor.\ Mat.\ Fiz.\ {\bf 25} (1975) 164.
\bibitem{town1} J.P.~Gauntlett, J.~Gomis, P.K.~Townsend,
{\it Particle actions as Wess-Zumino terms for spacetime (super)symmetry groups},
Phys.\ Lett.\ B {\bf 249} (1990) 255.
\bibitem{town2} J.P.~Gauntlett, K.~Itoh, P.K.~Townsend,
{\it Superparticle with extrinsic curvature},
Phys.\ Lett.\ B {\bf 238} (1990) 65.
\bibitem{GKP} J.~Gomis, K.~Kamimura, J.M.~Pons,
{\it Non-linear realizations, Goldstone bosons of broken Lorentz rotations and effective actions for p-branes},
Nucl.\ Phys.\ B {\bf 871} (2013) 420, arXiv:1205.1385[hep-th].
\bibitem{GKW} J.~Gomis, K.~Kamimura, P.C.~West,
{\it The construction of brane and superbrane actions using non-linear realisations},
Class.\ Quant.\ Grav.\ {\bf 23} (2006) 7369, arXiv:hep-th/0607057.
\bibitem{asluk1} J.A.~de~Azc\'{a}rraga, J.~Lukierski,
{\it Supersymmetric particles with internal symmetries and central charges},\\
Phys.\ Lett.\ B {\bf 113} (1982) 170.
\bibitem{ksquare} M.~Pav\v{s}i\v{c},
{\it Classical motion of membranes, strings and point particles with extrinsic curvature},\\
Phys.\ Lett.\ B {\bf 205} (1988) 231;\\
{\it The quantization of a point particle with extrinsic curvature leads to the Dirac equation},\\
Phys.\ Lett.\ B {\bf 221} (1989) 264.

A.A.~Kapustnikov, A.~Pashnev, A.~Pichugin,
{\it The canonical quantization of the kink model beyond the static solution},
Phys.\ Rev.\ D {\bf 55} (1997) 2257, arXiv:hep-th/9608124.
\bibitem{rigid} R.D.~Pisarski,
{\it Theory of curved paths},
Phys.\ Rev.\ D {\bf 34} (1986) 670.

M.S.~Plyushchay,
{\it Massive relativistic point particle with rigidity},
Int.\ J.\ Mod.\ Phys.\ A {\bf 4} (1989) 3851;\\
{\it Relativistic model of anyon,}
Phys.\ Lett.\ B {\bf 248} (1990) 107.

V.V.~Nesterenko,
{\it The singular lagrangians with higher derivatives},
J.\ Phys.\ A {\bf 22} (1989) 1673;\\
{\it Relativistic particle with curvature in an external electromagnetic field},
Int.\ J.\ Mod.\ Phys.\ A {\bf 6} (1991) 3989.

J.~Isberg, U.~Lindstr\"{o}m, H.~Nordstr\"{o}m, J.~Grundberg,
{\it Canonical quantization of a rigid particle},\\
Mod.\ Phys.\ Lett.\ A {\bf 5} (1990) 2491.

A.~Deriglazov, A.~Nersessian,
{\it Rigid particle revisited: extrinsic curvature yields the Dirac equation},\\
arXiv:1303.0483[hep-th].
\bibitem{anyon} R.~Jackiw, V.P.~Nair,
{\it Relativistic wave equations for anyons},
Phys.\ Rev.\ D {\bf 43} (1991) 1933.

C.-H.~Chou, V.P.~Nair, A.P.~Polychronakos,
{\it On the electromagnetic interactions of anyons},\\
Phys.\ Lett.\ B {\bf 304} (1993) 105, arXiv:hep-th/9301037.

J.L.~Cortes, M.S.~Plyushchay, {\it Anyons as spinning particles},\\
Int.\ J.\ Mod.\ Phys.\ A {\bf 11} (1996) 3331, arXiv:hep-th/9505117.

I.V.~Gorbunov, S.M.~Kuzenko, S.L.~Lyakhovich,
{\it On the minimal model of anyons},\\
Int.\ J.\ Mod.\ Phys.\ A {\bf 12} (1997) 4199, arXiv:hep-th/9607114.

I.V.~Gorbunov, S.M.~Kuzenko, S.L.~Lyakhovich,
{\it $N{=}1, D{=}3$ superanyons, $osp(2|2)$ and the deformed Heisenberg algebra},
Phys.\ Rev.\ D {\bf 56} (1997) 3744,  arXiv:hep-th/9702017.

A.~Nersessian, {\it On the geometry of relativistic anyon},\\
Mod.\ Phys.\ Lett.\ A {\bf 12} (1997) 1783, arXiv:hep-th/9704182.
\bibitem{asluk2} J.A.~de~Azc\'{a}rraga, J.P.~Gauntlett, J.M.~Izquierdo, P.K.~Townsend,
{\it Topological extensions of the supersymmetry algebra for extended objects},
Phys.\ Rev.\ Lett.\ {\bf 63} (1989) 2443.
\bibitem{Iv} S.~Bellucci, E.~Ivanov, S.~Krivonos,
{\it Superbranes and super Born-Infeld theories from nonlinear realizations},\\
Nucl.\ Phys.\ Proc.\ Suppl.\ {\bf 102} (2001) 26, arXiv:hep-th/0103136.
\bibitem{BG2} J.~Bagger, A.~Galperin,
{\it New Goldstone multiplet for partially broken supersymmetry},\\
Phys.\ Rev.\ D {\bf 55} (1997) 1091, arXiv:hep-th/9608177.
\bibitem{RT} M.~Ro\v{c}ek, A.A.~Tseytlin,
{\it Partial breaking of global $D{=}4$ supersymmetry, constrained superfields, and 3-brane actions},
Phys.\ Rev.\ D {\bf 59} (1999) 106001, arXiv:hep-th/9811232.
\bibitem{IK1} E.~Ivanov, S.~Krivonos,
{\it $N{=}1, D{=}4$ supermembrane in the coset approach},\\
Phys.\ Lett.\ B {\bf 453} (1999) 237, arXiv:hep-th/9901003.
\bibitem{BKS1} S.~Bellucci, S.~Krivonos, A.~Sutulin,
{\it Supersymmetric component actions via coset approach},\\
Phys.\ Lett.\ B {\bf 726} (2013) 497, arXiv:1306.1115[hep-th].
\bibitem{BKKS1} S.~Bellucci, N.~Kozyrev, S.~Krivonos, A.~Sutulin,\\
{\it Partial breaking of global supersymmetry and super particle actions},
arXiv1309.3902[hep-th].
\bibitem{VA} D.V.~Volkov, V.P.~Akulov,
{\it Possible universal neutrino interaction},
JETP Lett.\ {\bf 16} (1972) 438;\\
{\it Is the neutrino a Goldstone particle?}
Phys.\ Lett.\ B {\bf 46} (1973) 109.
\bibitem{tmp98} A.~Nersessian,
{\it The Hamiltonian formalism for the generalized rigid particles},\\
Theor.\ Math.\ Phys.\ {\bf 117} (1998) 1214, arXiv:hep-th/9805009.
\bibitem{plyushchay} M.S.~Plyushchay,
{\it Relativistic particle with torsion, Majorana equation and fractional spin},\\
Phys.\ Lett.\ B {\bf 262} (1991) 71;\\
{\it The model of relativistic particle with torsion,}
Nucl.\ Phys.\ B {\bf 362} (1991) 54.
\bibitem{ramos}
A.~Nersessian, E.~Ramos,
{\it Massive spinning particles and the geometry of null curves},\\
Phys.\ Lett.\ B {\bf 445} (1998) 123, arXiv:hep-th/9807143;\\
{\it A Geometrical particle model for anyons,}
Mod.\ Phys.\ Lett.\ A {\bf 14} (1999) 2033, arXiv:hep-th/9812077.
\end{thebibliography}
\end{document}